\documentclass[12pt]{article}
\usepackage{amsmath,amssymb,amsfonts}
\usepackage{bm}
\usepackage{multirow}
\usepackage{url}
\usepackage{epsfig}
\usepackage[paper=letterpaper,margin=1.0in]{geometry}
\usepackage{comment}

\newcommand{\be}{\begin{equation}}
\newcommand{\ee}{\end{equation}}
\newcommand{\bea}{\begin{eqnarray}}
\newcommand{\eea}{\end{eqnarray}}
\newcommand{\ba}{\begin{array}}
\newcommand{\ea}{\end{array}}
\newcommand{\nn}{\nonumber}
\newcommand{\tr}{\mathrm{tr}\,}
\newcommand{\IC}{\mathbb{C}}
\newcommand{\IP}{\mathbb{P}}

\newcommand{\cM}{{\cal M}}

\newcommand{\cF}{{\cal F}}

\def\eps{\epsilon^{\alpha \beta}}
\def\barH{\overline{H}}

\begin{document}

\thispagestyle{empty}

\renewcommand{\thefootnote}{\fnsymbol{footnote}}

${}$
\vskip2cm
\begin{center}

{\Large \textbf{Testing R-parity with Geometry}}

\vskip1cm

\textbf{
Yang-Hui He${}^{a,b,c,}$\footnote{\texttt{hey@maths.ox.ac.uk}},
Vishnu Jejjala${}^{d,}$\footnote{\texttt{vishnu@neo.phys.wits.ac.za}},
Cyril Matti${}^{a,d,}$\footnote{\texttt{cyril.matti.1@city.ac.uk}},
Brent D.\ Nelson${}^{e,}$\footnote{\texttt{b.nelson@neu.edu}}}

\vskip0.5cm

${}^a$\textit{Department of Mathematics, City University, London,\\
Northampton Square, London EC1V 0HB, UK}
\vskip0.25cm

${}^b$\textit{School of Physics, NanKai University, Tianjin, 300071, P.R.~China}
\vskip0.25cm

${}^c$\textit{Merton College, University of Oxford, OX1 4JD, UK}
\vskip0.25cm

${}^d$\textit{Mandelstam Institute for Theoretical Physics, NITheP, and School of Physics,\\
University of the Witwatersrand, Johannesburg, WITS 2050, South Africa}
\vskip0.25cm

${}^e$\textit{Department of Physics, Northeastern University, Boston, MA 02115, USA}

\end{center}

\vskip0.5cm

\begin{abstract}
  We present a complete classification of the vacuum geometries of all renormalizable superpotentials built from the fields of the electroweak sector of the MSSM.
  In addition to the Severi and affine Calabi-Yau varieties previously found, new vacuum manifolds are identified; we thereby investigate the geometrical implication of theories which display a manifest matter parity (or R-parity) via the distinction between leptonic and Higgs doublets, and of the lepton number assignment of the right-handed neutrino fields.

  We find that the traditional R-parity assignments of the MSSM more readily accommodate the neutrino see-saw mechanism with non-trivial geometry than those superpotentials that violate R-parity. However there appears to be no geometrical preference for a fundamental Higgs bilinear in the superpotential, with operators that violate lepton number, such as $\nu H \barH$, generating vacuum moduli spaces equivalent to those with a fundamental bilinear.
\end{abstract}

\newpage

\renewcommand{\thepage}{\arabic{page}}
\setcounter{page}{1}
\def\thefootnote{\arabic{footnote}}
\setcounter{footnote}{0}


\section{Introduction}\label{Introduction}

The vacuum of an ${\cal N}=1$ supersymmetric quantum field theory consists of field configurations that satisfy the F-term and D-term constraints. Recent advances in understanding the vacuum moduli space of the minimal supersymmetric standard model (MSSM) establish that non-trivial geometrical characteristics correlate to specific parameter choices of the theory such as the number of generations of matter fields, the number of pairs of Higgs doublets, and the vanishing of coupling constants~\cite{Gray:2005sr,Gray:2006jb,He:2014loa,He:2014oha}. These are the first hints that a \textit{bottom up} strategy to studying low dimensional effective field theories might yield information about the structure of its higher energy completion~\cite{Douglas:1996sw}. Frequently, the vacuum geometries of semi-realistic quantum field theories have interesting mathematical properties. They may, for instance, be Calabi--Yau, as happens for orbifolds of $\mathbb{C}^3$ by discrete subgroups of $SU(2)$ and $SU(3)$~\cite{Benvenuti:2006qr} or for supersymmetric QCD (SQCD)~\cite{Gray:2008yu}.

The geometrical nature of the vacuum space is intimately linked with phenomenology, given its relevance and importance for understanding patterns of supersymmetry breaking and for model building. Nevertheless, the full moduli space of the MSSM remains unknown. This is due to the computational complexity involved in deducing relations among the nearly one~thousand holomorphic gauge invariant operators (GIOs) needed to parameterize the vacuum as an algebraic variety. However, much is already known about the electroweak sector, namely the specific locus in the complete moduli space where the vacuum expectation values of fields charged under $SU(3)$ vanish. This ensures an unbroken color symmetry at low energies. We report in~\cite{He:2014oha} that phenomenologically viable theories prefer a geometry described by Severi varieties. These are algebraic varieties whose secant variety is not a projective space, and only four such examples exist.\footnote{This is connected to the existence of four division algebras: the real numbers, the complex numbers, the quaternions, and the octonions.} The appearance of this rare feature may suggest that it has phenomenological significance.

In this paper, we seek to establish how special these Severi varieties are in the context of electroweak theories. That is, we consider all possible superpotentials with renormalizable terms, irrespective of whether they are phenomenologically plausible, and study the corresponding vacuum structures. This comparison of theories enables us to isolate properties associated to the more phenomenologically interesting ones.

One phenomenological feature of particular interest for us will be the presence of a $\mathbb{Z}_2$ matter-parity, sometimes referred to as R-parity, in the structure of superpotential interactions. This multiplicatively-conserved quantum number is typically defined as
\begin{equation} P_R = (-1)^{3(B-L)+2s}\, , \label{RP} \end{equation}
where $B$ and $L$ are the baryon and lepton numbers, respectively, of the fields in a supermultiplet, and $s$ represents the spin of the individual components of the supermultiplet. When discussing the allowed terms in the superpotential, it is common to assign a $P_R$ value to the whole supermultiplet via the lowest component, thus using $s=0$ in~(\ref{RP}). Such a parity, when present, is effective at preventing rapid proton decay and may allow for a natural supersymmetric dark matter candidate~\cite{Barbier:2004ez}. It is thus of great moment to ask whether geometrical techniques can shed light on the nature of any such R-parity.

In the context of the current analysis, $P_R$ effectively amounts to the assignment of definite lepton number to the supermultiplets, which in turn means asking whether there is any distinction between lepton and Higgs doublets, from a geometrical point of view. It also gets at the thorny question of the nature of neutrinos in string contexts, and the manner by which they receive masses in the MSSM~\cite{Langacker:2011bi}. In the current work we will address the emergence of a lepton-Higgs distinction in the vacuum moduli spaces of electroweak gauge theories, and correlate geometrical properties with the traditional R-parity assignments assumed in the construction of the MSSM.

In performing this analysis, we apply a well known algorithm for computing the vacuum moduli space. Recall that for a general ${\cal N}=1$ globally supersymmetric theory in four dimensions, we have the action
\bea
S &=& \int d^4x\ \Bigg[ \int d^4\theta\ \Phi_i^\dagger e^V \Phi_i \;+ 
\Bigg(\frac{1}{4g^2} \int d^2\theta\ \tr{W_\alpha W^\alpha} + \int d^2\theta\ W(\Phi) + {\rm h.c.} \Bigg) \Bigg]\;,
\eea
where $\Phi_i$ are chiral superfields, $V$ is a vector superfield, $W_\alpha$ are chiral spinor superfields and the superpotential $W$ is a holomorphic function of the superfields $\Phi_i$. The chiral spinor superfields are given by the gauge field strength $W_\alpha = i\overline{D}^2 e^{-V} D_\alpha e^V$. The vacuum of such a theory consists of the vacuum expectation values $\phi_{i 0}$ of the scalar components of the superfields $\Phi_i$ that simultaneously solve the F-term equations,
\be
\label{fterm}
\left. F^i=\frac{\partial W(\phi)}{\partial \phi_i}
\right|_{\phi_i=\phi_{i 0}} = 0  \; ,
\ee
and the D-term equations, which in Wess--Zumino gauge read
\be
\label{dterm}
D^A = \sum_i \phi_{i 0}^\dagger\, T^A\, \phi_{i 0} = 0 ~,
\ee
where $T^A$ are generators of the gauge group in the adjoint representation. Solutions to the above equations describe the vacuum moduli space $\cal M$ as an algebraic variety in the fields $\phi_{i 0}$.

While the F-term equations~\eqref{fterm} simply correspond to a Jacobian and are straightforward to solve, it tends to be much more difficult to satisfy the D-term equations~\eqref{dterm}. Nonetheless~\cite{Gray:2009fy, Mehta:2012wk} demonstrates that solving~\eqref{fterm} and~\eqref{dterm} is equivalent to an elimination problem (see also~\cite{He:2014loa} for a summary). For every solution to the F-term equations, there exists a unique solution to the D-terms in the completion of the complexified orbit of the gauge group~\cite{Buccella:1982nx,Procesi:1985hr,Gatto:1986bt,Witten:1993yc,Luty:1995sd}. Such orbits are specified by the minimal set of holomorphic gauge invariant operators in the theory, and it is therefore sufficient to consider the description of the moduli space as the symplectic quotient of the space of F-flat field configurations by the complexified gauge group.\footnote{Strictly speaking, ${\cal M} = {\cal F}/\!/G^C$ is a geometric invariant theory (GIT) quotient.}
Algebraically, this corresponds to the ring map $D$ from the quotient ring $\cF=\IC[\phi_1, \ldots, \phi_n] \left/ \left\langle\frac{\partial {W}}{\partial \phi_i}\right\rangle\right.$ to the ring $S = \IC[y_{j=1,\ldots,k}]$,
\be
\label{map}
\cM \simeq {\rm Im}\left(\cF\quad \stackrel{D=\{r_j( \{ \phi_i \})\}}{\xrightarrow{\hspace*{2cm}}} \quad S \right)\;,
\ee
where $r_j( \{ \phi_i \})$ are polynomials in the scalar fields $\phi_i$ corresponding to the GIOs of the theory. Here, $j=1,\ldots,k$ with $k$ the total number of GIOs. We employ the computational algebraic geometry packages \texttt{Macaulay~2}~\cite{mac} and \texttt{Singular}~\cite{sing} to solve the elimination problem. The coupling constants (or equivalently the GIO content) of the superpotential $W$ thus determines the vacuum moduli space. The goal of this paper is to characterize the vacuum geometry with respect to choices of $W$.

The paper is organized as follows.
In Section~\ref{GIO_sec}, we present the structure of the electroweak GIOs and the related algebraic varieties defined by their relations and syzygies.
The F-term equations represent additional constraints on these varieties and therefore lead to subvarieties.
In Section~\ref{Fterm_sec}, we classify the subvarieties according to all possible choices of superpotentials without right handed neutrinos.
In Section~\ref{Neut_sec}, we study the effect of introducing right handed neutrinos to the theory.
Finally in Section~\ref{Discussion_sec}, we summarize the results of our investigations and compare the various models. We make some initial observations about the geometrical significance of R-parity and the nature of the neutrino multiplet, while outlining some future research directions inspired from these observations.


\section{Gauge Invariant Operators and Syzygies}\label{GIO_sec}

In the present work we restrict our attention to the fields that constitute the electroweak sector of the MSSM. In this particular section that will mean the fields $L^i_\alpha$, $e^i$, $H_\alpha$, and $\barH_\alpha$: the lepton doublet and singlet charged lepton, the up-type Higgs doublet, and the down-type Higgs doublet, respectively. The lepton and the electron have a flavor index $i=1,2,3$. As they transform as doublets of $SU(2)$, the lepton doublet and the up-type and down-type Higgs fields carry an $SU(2)$ index $\alpha=1,2$. All other MSSM matter fields are endowed with a color index and thus have vanishing vacuum expectation values. For now, we do not include the right handed neutrinos $\nu^i$, which are uncharged under the Standard Model gauge group. As such, we have $13$ component fields in the electroweak sector, which means that there is {\em a priori} as many F-term equations arising from derivatives of the superpotential. A minimal list of generators for the GIOs of the MSSM electroweak sector is given in Table~\ref{gio-ew}.

\begin{table}[t]
{\begin{center}
\begin{tabular}{|c||c|c|c|}\hline
\mbox{Type} & \mbox{Explicit Sum} & \mbox{Index} & \mbox{Number} \\
\hline \hline
$LH$  & $L^i_\alpha H_\beta \eps$ & $i=1,2,3$ & $3$ \\ \hline
$H\barH$ & $H_\alpha \barH_\beta \eps$ & & 1  \\ \hline
$LLe$ & $L^i_\alpha L^j_\beta e^k \eps$ & $i,k=1,2,3;
j=1,2$ & $9$  \\ \hline
$L\barH e$ & $L^i_\alpha \barH_\beta \eps e^j$ & $i,j=1,2,3$ & $9$ \\
\hline
\end{tabular}
\end{center}}{\caption{\label{gio-ew}{\bf Generators of the GIOs for the electroweak sector of the MSSM.}}}
\end{table}

Since the vacuum geometry lives in the image of the ring map~\eqref{map}, the F-term constraints impose restrictions on the ideal of relations among the GIOs. These constraints have two effects:
\begin{enumerate}
\item[i.] they might simply lift some GIOs from the vacuum, by forcing these GIOs to have vanishing expectation values;
\item[ii.] they might introduce extra internal relations among the GIOs.
\end{enumerate}

The geometry, or geometries plural, defined solely by the GIOs corresponds to the ideal of their relations, or syzygies.\footnote{
The term \textit{syzygy}, borrowed from celestial mechanics, refers to relations between generators of a module.}
Such geometries serve, in some sense, as master spaces common to many of the outcomes we will study in this paper. These geometries may be thought of as the vacuum moduli spaces given by the non-vanishing GIOs when the F-term equations do not impose additional relations among themselves. This is the case when the effect of F-term equations is simply to lift some of the GIOs (outcome~i. above). It is therefore important to understand what are the possible varieties arising in this way.

First, let us consider the simplest cases. We begin with the set ${\cal S} = \{ LH, H\barH \}$,  which consists of four objects (note that we are suppressing the generation indices in our notation). We observe that these are freely generated: there are no relations among the generators. The geometry described is thus trivial and corresponds to $\IC^4$. This holds as well for any subset of these operators, in which case the geometry is given by $\IC^n$, where $n$ is the number of elements in ${\cal S}$.

The next-simplest case is the first for which non-trivial relations arise. We consider the nine objects defined by the set $\{ LLe\}$ or the nine objects defined by the set $\{ L\barH e\}$. Both of these types of generators can be written as a product of two $SU(2)$ singlets with opposite hypercharges:
\be
\chi^i_{1} \chi^j_{-1} \quad {\rm with} \quad i,j = 1,\dots 3 \; ,
\ee
where $\chi_{-1}$ stands for the electron $e$ (hypercharge $+1$) and $\chi_{1}$ is either $LL$ or $L\barH$ (both with hypercharge $-1$). It is straightforward to realize that, considering each type individually, these GIOs are subject to relations,
\be\label{Segre-rel}
(\chi^i_{1} \chi^j_{-1})(\chi^k_{1} \chi^l_{-1}) = (\chi^i_{1} \chi^l_{-1})(\chi^k_{1} \chi^j_{-1}) \; .
\ee
These equations define the Segre embedding of $\IP^2 \times \IP^2$ into $\IP^8$, where $\IP^n$ is the $n$-dimensional complex projective space. Indeed, each of the three $\chi_{\pm 1}$ gives rise to a coordinate in $\IP^2$. The operator $LL$ is described by a Grassmannian $Gr(3,2) = \IP^2$ due to the antisymmetrization of the indices, and $L\barH$ and $e$ both have a free index running from $1$ to $3$ and therefore corresponds to a $\IP^2$ upon projectivization. The embedding therefore is the following. Take $[x_0:x_1:x_2]$ and $[z_0:z_1:z_2]$ as the homogeneous coordinates of two $\IP^2$s and consider the quadratic map
\begin{equation} \label{Segre}
\begin{array}{ccccc}
LL {\rm ~or~} L\barH&  & e &   & LLe {\rm ~or~} L\barH e \\
\IP^2 & \times & \IP^2 & \longrightarrow & \IP^8 \\
\mbox{$[x_0 : x_1 : x_2]$}
& &  
\mbox{$[z_0 : z_1 : z_2]$}
& \mapsto & x_i z_j \\
\end{array} ~,
\end{equation}
where $i,j=0,1,2$, giving precisely $3^2 = 9$ homogeneous coordinates of $\IP^8$. We can summarize this space with the standard notation $(8|5,6|2^9)$, where $(k | d, \delta | m_1^{n_1} m_2^{n_2} \ldots )$ means an affine variety of complex dimension $d$, realized as an affine cone over a projective variety of dimension $d-1$ and degree $\delta$, given as the intersection of $n_i$ polynomials of degree $m_i$ in $\IP^k$.

The Segre embedding is a Severi variety and its corresponding Hilbert series of the first kind is
\be\label{H_EW}
\frac{1+4t+t^2}{(1-t)^5} \ .
\ee
Since the numerator has coefficients that are palindromic, the space is an affine Calabi--Yau manifold, in the sense that it has a trivial canonical sheaf.\footnote{ A more precise statement is the following. By Stanley's theorem, the numerator of a Hilbert series of a graded Cohen--Macaulay domain $R$ is palindromic if and only if $R$ is Gorenstein~\cite{stanley}. For affine varieties, the Gorenstein condition is equivalent to the Calabi--Yau condition~\cite{bh}. See also~\cite{He:2014oha,Gray:2008yu}.}
The exponent in the denominator gives the dimension of the variety.

One important point to emphasize is the fact that $LLe$ and $L\barH e$ give rise to the same varieties. This stems from the fact that we have three $LL$ combinations as well as three $L\barH$ combinations. This coincidence is due to the number of particle generations and Higgses. From the anti-symmetrization of the $SU(2)$ indices in $LL$, there are only three possible combinations: $\binom{3}{2}=3$. The $LL$ therefore yields a Grassmannian $Gr(3,2)$, but we have $Gr(3,2)=\IP^2$. On the other hand, the flavor index is free on $L\barH$, but, as the Higgs field bears no index, there are also only three $L\barH$. The ``symmetry'' between $LLe$ and $L\barH e$ thus only holds for three generations of matter fields and one generation of Higgs fields. Thus far, vacua based solely on relations amongst GIOs fail to distinguish between leptons and Higgs fields.

Let us now consider the two sets together ${\cal S}' = \{LLe , L\barH e \}$. Extra relations among the 18~objects in this set will now appear due to the fact that both types share common fields. With the help of~\texttt{Macaulay~2}, we find that these relations describe a seven-dimensional variety given by $51$ quadratics $( 17 | 7 , 30 | 2^{51} )$. The Hilbert series obtained is
\be\label{LLeLHIe}
\frac{1+11t+15t^2+3t^3}{(1-t)^7} \ .
\ee
Thus, because the numerator does not have palindromic coefficients, the geometry is no longer Calabi--Yau.

We can understand this variety in the following manner. Since $L$ and $\barH$ bear the same quantum numbers, they are not distinguishable from a gauge theory standpoint. Each GIO in the set ${\cal S}'$ is thus a product $\chi_1\chi_{-1}$, where $\chi_{-1}$ are the electron $e$ and $\chi_1$ are the $\binom{4}{2} = 6$ possible field configurations $\phi_\alpha\phi_\beta\eps$ with $\phi \in \{L,\barH\}$. This corresponds to the Grassmannian $Gr(4,2)$. Therefore, we have the following embedding originating from the relations~\eqref{Segre-rel}:
\begin{equation}\label{Mattifold_4}
\begin{array}{ccccc}
\{LL,L\barH\} &  & e &  & \{LLe, L\barH e\} \\
\mbox{$Gr(4,2)$} & \times & \IP^2 & \longrightarrow & \IP^{17} \\
\mbox{$[x_0 : x_1 : x_2 : x_3 : x_4 : x_5]$}
& &  
\mbox{$[z_0 : z_1 : z_2]$}
& \mapsto & x_i z_j \\
\end{array} ~,
\end{equation}
where $x_i$ are Pl\"ucker coordinates subject to the relation,
\begin{equation}
x_0x_5 = x_1x_4 - x_2x_3 \ .
\end{equation}
The dimension of $Gr(4,2)$ is four, as we have $\dim Gr(n,r) = r(n-r)$. Therefore, with the three additional electron coordinates, we obtain a seven dimensional variety as required.

Finally, we should compute the geometry when all possible generators of Table~\ref{gio-ew} are considered together, ${\cal S}'' = \{LLe, L\barH e, LH, H\barH\}$. We obtain an irreducible nine dimensional variety given by $63$ quadratic relations: $( 21 | 9 , 56 | 2^{63} )$. The Hilbert series is
\be\label{BigCYHS}
\frac{1+13t+28t^2+13t^3+t^4}{(1-t)^9} \ .
\ee
Remarkably, the variety is a Calabi--Yau again. This of course corresponds to the vacuum geometry for the MSSM electroweak theory when there is no superpotential: $W=0$.

The geometry can be explained in the following way.
The set of ten operators 
\begin{equation}
 \{LL, L\barH, LH, H\barH\} \; ,
\end{equation}
describes a Grassmannian $Gr(5,2)$ from the antisymmetrized product of the $SU(2)$ indices.
This is a six dimensional variety.
The three electron fields give three additional dimensions and the resulting total variety is therefore nine dimensional:
\begin{equation}
\begin{array}{ccccc}
\{LL, L\barH, LH, H\barH\}&& e & & \{LLe, L\barH e, LH, H\barH\} \\
\mbox{$Gr(5,2)$} & \times & \IP^2 & \longrightarrow & \IC^{22} \\
\mbox{$[x_0 : x_1 : x_2 : x_3 : x_4 : x_5 : x_6 : x_7 : x_8 : x_9]$}
& &  
\mbox{$[z_0 : z_1 : z_2]$}
& \rightarrow & (x_i z_j , x_6 , x_7 , x_8 , x_9) \\
\end{array} ~,\label{BigCY}
\end{equation}
where $i = 0, \ldots 5$ and $j= 0, \ldots 2$ and the Grassmannian coordinates $x$ are subject to the relations:
\bea
\nn
x_0x_5+x_2x_3-x_1x_4=0 \; ,\\
\nn
x_0x_8+x_2x_6-x_1x_7=0\; ,\\
\nn
x_0x_9+x_4x_6-x_3x_7=0\; ,\\
\nn
x_1x_9+x_5x_6-x_3x_8=0\; ,\\
x_2x_9+x_5x_7-x_4x_8=0\; .
\eea
In~\eqref{BigCY}, the $x_0, \ldots, x_5$ variables corresponding to $\{LL, L\barH\}$ are contracted with an electron field $z_j$, while the remaining four variables $x_6, \ldots, x_9$ correspond to the hypercharge neutral operators $\{LH, H\barH\}$. In this case the embedding space is not a projective space.

We are now ready to understand the effects of F-term equations on the vacuum geometry resulting from the sets of GIOs considered. This is developed in the next section.


\section{F-terms Constraints}\label{Fterm_sec}

In addition to the syzygies of the GIOs, the vacuum geometry is governed by the constraints arising from F-term equations. The most general superpotential consistent with R-parity at the renormalisable level that can be written with GIOs of Table~\ref{gio-ew} corresponds to
\bea
W_{\rm minimal} &=& \; C^0 \sum_{\alpha, \beta} H_\alpha \barH_\beta \eps  + \label{renorm} \sum_{i,j}C^3_{ij} e^i \sum_{\alpha, \beta} L^j_{\alpha} \barH_\beta \eps ~,
\eea
where $C^0$ and $C^3$ are coupling constants. This is the electroweak superpotential of the MSSM (without neutrinos). The imposition of R-parity, in the present context, amounts to requiring that each term in~$W$ contain an even number of multiplets which carry lepton number. In search of a selection mechanism based on geometrical principles, we would like to understand what, if anything, is special about the above superpotential from the perspective of vacuum geometry. For this we will investigate deformations to~(\ref{renorm}), considering combinations of the electroweak GIOs, without necessarily insisting on conserved R-parity. We can then compare the corresponding vacuum geometry for each superpotential.

In the absence of right-handed neutrinos, the allowed supertpotential terms are also the GIOs listed in Table~\ref{gio-ew}. They can be grouped naturally into two categories. First, we consider the set,
\be
{\cal T} = \left\{L\barH e, LLe \right\} \; , \label{setT}
\ee
which we will refer to as {\bf elemental trilinears}. This terminology comes from the fact that they are the product of three fields (trilinear) that do not constitute a product of other GIOs. Thus they are elemental, as opposite to composite.
In addition to these trilinears, we need to consider the elemental bilinears,
\be
{\cal B} = \left\{H\barH, LH\right\} \; . \label{setB}
\ee
Again, they are products of two fields which are not GIOs themselves, hence the name elemental bilinears.

For the sake of generality, we will consider non-vanishing random coupling coefficients and will ask which combinations of the above operators give rise to a non-trivial variety -- that is varieties which are different from a set of points and planes, {\em e.g.} a non-trivial Hilbert series -- when included in the superpotential. The possible superpotential combinations can be divided into three categories. In ascending order of complexity, these are the case containing two elements of $\cal T$, the case containing no element of $\cal T$, and the cases with only one element of $\cal T$. In the next sections, we will consider these four options in turn.

\subsection{Cases with two elemental trilinears}
\label{sec:31}

\begin{table}[ht]
{\begin{center}
\begin{tabular}{|c|c|c||c|c|c|}\hline
\multicolumn{3}{|c||}{$W$ \mbox{terms}} & \multicolumn{3}{c|}{\mbox{Vacuum moduli space}}\\ \hline
$LLe + L\barH e$ & $LH$ & $H\barH$ & \mbox{dim} &  \mbox{geometry} &  \mbox{operators}
\\
\hline \hline
\checkmark&&& 4 & $\IC^4$& $LH, H\barH$\\ \hline
\checkmark&\checkmark&& 0 & \mbox{point at the origin} & - \\ \hline
\checkmark&&\checkmark& 0 & \mbox{point at the origin} & - \\ \hline
\checkmark&\checkmark&\checkmark& 0 & \mbox{point at the origin} & - \\ \hline
\end{tabular}
\end{center}}{\caption{\label{two-trilin}{\bf Vacuum moduli space for superpotentials $W$ including two elemental trilinears.} The dimension (dim), a description of the geometry, and which operators are non-vanishing in the vacuum are presented against the GIOs present in the superpotential (marked with a tick).}}
\end{table}

The first thing to note is that non-trivial geometry can only arise from relations between the $LLe$ and/or $L\barH e$ operators, as described in Section~\ref{GIO_sec}. Including either term in the superpotential lifts the corresponding GIO. This is clear from the requirement that the F-term expressions arising from the singlet lepton vanish in the vacuum: $\partial W/\partial e^i = 0$. Thus, inclusion of both trilinears simultaneously in the superpotential can only lead to a trivial geometry (consisting of points or planes). When $W=LLe + L\barH e$ the vacuum is determined exclusively by the GIOs in the set ${\cal S} = \{ LH, H\barH \}$, so the vacuum geometry is $\IC^4$. Further adding elemental bilinears from~(\ref{setB}) introduces the up-type Higgs field $H$, whose F-term equation immediately lifts the bilinear GIOs from the vacuum as well. The four possible combinations of superpotentials are summarized in Table~\ref{two-trilin}.

\subsection{Cases without any elemental trilinear}
\label{sec:32}

\begin{table}[ht]
{\begin{center}
\begin{tabular}{|c|c||c|c|c|c|}\hline
\multicolumn{2}{|c||}{$W$ \mbox{terms}} & \multicolumn{4}{c|}{\mbox{Vacuum moduli space}}\\ \hline
$LH$ & $H\barH$ & \mbox{description} & \mbox{Hilbert series}  & \mbox{geometry}& \mbox{operators}
\\
\hline \hline
\checkmark&& $( 8 | 5 , 6 | 2^9 )$&${(1+4t+t^2)}/{(1-t)^5}$&\mbox{Segre} &$LLe, L\barH e$\\ \hline
&\checkmark& $( 8 | 5 , 6 | 2^9 )$&${(1+4t+t^2)}/{(1-t)^5}$&\mbox{Segre} &$LLe$\\ \hline
\checkmark&\checkmark& $( 8 | 5 , 6 | 2^9 )$&${(1+4t+t^2)}/{(1-t)^5}$&\mbox{Segre} &$LLe, L\barH e$\\ \hline
\end{tabular}
\end{center}}{\caption{\label{no-trilin}{\bf Vacuum moduli space for superpotentials $W$ without any elemental trilinear.} The description according to our standard notation, the Hilbert series, the geometry, and which operators are non-vanishing in the vacuum are presented against the GIOs present in the superpotential (marked with a tick).}}
\end{table}

When considering superpotentials without any trilinear, we obtain the Segre variety as the vacuum geometry for all three combinations of $LH$ and $H\barH$. Indeed, it is not possible to lift $LLe$ or $L\barH e$ with $LH$ and $H\barH$ terms only. The $LH$ and $H\barH$ GIOs are lifted from the vacuum either by the $F_L$-term and/or the $F_{\barH}$-term equations, both of which impose the vanishing of the $H$-field in the vacuum for the three superpotential cases. 

The vacuum geometry is therefore obtained from the embedding~\eqref{Mattifold_4} with the remaining $F_H$-term equation imposed on it. Two distinct cases occur depending on whether $LH$ is present in $W$ or not. For the superpotential with $H\barH$ only, the fields $\barH$ must vanish from the $F_H$-term equation and thus the vacuum geometry is described by the relations among the $LLe$, that is the Segre embedding~\eqref{Segre}.
For the other two possible superpotentials, the $F_H$-term equation imposes one linear constraint on the set of fields $\{L,\barH \}$. This linear constraint will reduce the Grassmannian $Gr(4,2)$ of the embedding~\eqref{Mattifold_4} down to a Grassmannian $Gr(3,2)$, which is equivalent to $\IP^2$. Thus, we obtain an embedding of the Segre type again. 
In other words, despite having both $LLe$ and $L\barH e$ in the vacuum, the F-term constraints impose the requirement that these two types of GIOs be related to each other. Surprisingly, we cannot obtain a vacuum with both $LLe$ and $L\barH e$ being fully unconstrained. Thus, we never obtain the geometry from the embedding~\eqref{Mattifold_4} as the vacuum geometry. We will revisit this circumstance in the presence of right-handed neutrino fields in Section~\ref{Neut_sec}. The three cases are summarised in Table~\ref{no-trilin}.

\subsection{Cases with only one elemental trilinear}

Let us now turn to the more complicated cases of superpotentials with one, and only one, trilinear from $\cal T$. We can easily observe that the trilinear present in the superpotential will be lifted from the vacuum due to the $F_e$-term equation. However, the geometry will be different for the two possible sets of superpotentials, and it is convenient to develop each case separately.

\subsubsection{$L\barH e$}
\label{secLHe}

Let us start with only one term -- the $L\barH e$ elemental trilinear -- in the superpotential:
\begin{equation}
\label{WLHIe}
W = C^3_{ij} \, L^i_\alpha \barH_\beta \, \eps \, e^j \; .
\end{equation}
Assuming a non-singular $C^3_{ij}$ coefficient matrix, it immediately follows from the $F_e$-term equations that $L\barH$, and hence $L\barH e$, must vanish. We also have three $L$-term equations,
\begin{equation}
\barH_\beta \, \eps \, e^j = 0 \; ,
\end{equation}
where we have used the inverse of the coefficient matrix $C^3_{ij}$. These constraints imply $H\barH e^i = 0$, which has two kind of solutions. Either $e^i \neq 0$, which implies $H\barH = 0$, or $e^i = 0$, which implies that $LLe = 0$ and $H\barH$ is unconstrained. In this latter case, $LH$ is also unconstrained, and we have a copy of $\IC^4$ given by $\{LH, H\barH\}$. The full vacuum geometry thus consists of two branches, a trivial branch $\IC^4$, and another non-trivial one including the operators $LLe$ and $LH$.

The variety described by the ideal of relations among the $LLe$ and $LH$ operators are given by the following embedding,
\begin{equation}
\begin{array}{ccccccc}
\{LL, LH\}&& e && \{LLe, LH, H\barH\}\\
\mbox{$Gr(4,2)$} & \times & \IP^2 & \longrightarrow & \IC^{13} \\
\mbox{$[x_0 : x_1 : x_2 : x_3 : x_4 : x_5]$} &
& 
\mbox{$[z_0 : z_1 : z_2]$}
& \rightarrow & (x_i z_j, x_3 , x_4 , x_5 ) \\
\end{array} ~,\label{LHLLetemp}
\end{equation}
where $i = 0, \ldots 2$ and $j= 0, \ldots 2$ and the Grassmann coordinates $x_i$ are subject to the Pl\"ucker relation
\begin{equation}
 x_0x_5 = x_1x_4 - x_2x_3 \ .
\end{equation}

We still have one condition left to satisfy from the F-term equation arising from the $\barH$-field
\begin{equation}\label{Hbar-term}
C^3_{ij} \, L^i_\alpha \, \eps \, e^j = 0 \; .
\end{equation}
Since the two indices $i,j$ are contracted, we cannot invert the coefficient matrix to simplify the equations. However, we can redefine the $e^i$ variables as
\begin{equation}
\tilde e_i \equiv C_{ij} e^j \; .
\end{equation}
Contracting~\eqref{Hbar-term} with $L^k_\beta$ we obtain a set of three equations
\begin{eqnarray} \label{simil1}
\nn
L^1_\beta (\tilde e_2L^2_\alpha \, \eps+ \tilde e_3L^3_\alpha \, \eps) &=&0 \; , \\
\nn
L^2_\beta (\tilde e_1 L^1_\alpha \, \eps + \tilde e_3L^3_\alpha \, \eps) &=&0 \; , \\\label{simil3}
L^3_\beta (\tilde e_1 L^1_\alpha \, \eps + \tilde e_2L^2_\alpha \, \eps) &=&0 \; .
\end{eqnarray}
This suggests that the embedding variables $z_i$ in~(\ref{LHLLetemp}) can be redefined in a related manner, such that the above conditions reduce to
\begin{equation}\label{xz}
x_i=\tilde z_i \; ,
\end{equation}
for $i=0\ldots 2$. This coordinate redefinition and identification was explored in greater detail in previous work~\cite{He:2014loa,He:2014oha}. In the absence of the extra $x_3,\ldots x_5$ variables (the LH GIOs) this would lead to the Veronese embedding of $\IP^2$ into $\IP^5$. Instead, we here obtain:
\begin{equation}
\begin{array}{ccc}
\{LL \sim \tilde e, LH\} && \{LL\tilde e, LH, H\barH\}\\
\mbox{$Gr(4,2)$} &\longrightarrow& \IC^{10} \\
\mbox{$[x_0 : x_1 : x_2 : x_3 : x_4 : x_5]$} &\rightarrow & (x_i x_j , x_3 , x_4 , x_5 , x_6) \\
\end{array} ~,\label{LHLLe}
\end{equation}
where $i,j = 0, \ldots 2$ and $i \leq j$ by symmetry. The notation $LLe \sim e$ indicates that here they are not independent degrees of freedom because there is a linear relation between them; we will use this notation henceforth. This embedding gives in fact the same Hilbert series as the Segre embedding and we will refer this vacuum moduli space for the superpotential~\eqref{WLHIe} as Segre~$\cup~\IC^4$.

Let us now consider deformations of the superpotential by including $LH$. These extra GIOs modify the $F_L$-term equations and introduce $F_H$-term equations. The effect of the $F_L$-term equations is to lift $LH$ and $H\barH$. Indeed, we have,
\begin{equation}
C^3_{ij} \, \barH_\beta \, \eps \, e^j +  C_i H_\beta \, \eps= 0 \; .
\end{equation}
Contracting back with $L^k_\alpha$ implies that $LH=0$, since the first term vanishes upon contraction by virtue of $L\barH =0$ from the $F_e$-term equation. Similarly, contracting with $\barH_\alpha$, we obtain $H\barH = 0$ as the first term vanishes from antisymmetrisation of the $\alpha$ and $\beta$ indices. Moreover, the new constraint $F_{H_\beta}=0$, contracted back with $L^j_\alpha$, gives
\begin{equation}
L^1_\alpha L^2_\beta \eps + L^1_\alpha L^3_\beta \eps =
L^1_\alpha L^2_\beta \eps - L^2_\alpha L^3_\beta \eps = L^1_\alpha L^3_\beta \eps + L^2_\alpha L^3_\beta \eps = 0 \; . \label{LLcond}
\end{equation}
This holds for the case when $H\barH$ is present in the superpotential and when it is not, since any additional terms in $F_{H_\beta}=0$ involving $\barH$ vanish upon contraction with $L$ by virtue of the $F_e$-term equations, as mentioned earlier. The conditions in~(\ref{LLcond}) imply that only one $LLe$ GIO is free, and the geometry becomes a trivial line $\IC$ with all other GIOs lifted.

The last case to consider is the superpotential with $L\barH e$ and $H\barH$. It corresponds to the MSSM and the vacuum geometry corresponds to the Segre variety. This has been discussed in great detail in~\cite{He:2014loa,He:2014oha}. The vacuum geometries encountered with the $L\barH e$ trilinear can thus be summarised as in Table~\ref{LbarHe-table}.

\begin{table}[htb]
{\begin{center}
\begin{tabular}{|c|c|c||c|c|c|c|}\hline
\multicolumn{3}{|c||}{$W$ \mbox{terms}} & \multicolumn{4}{c|}{\mbox{Vacuum moduli space}}\\ \hline
$L\barH e$ & $LH$ & $H\barH$ & \mbox{description} & \mbox{Hilbert series}  & \mbox{geometry}& \mbox{operators}
\\
\hline \hline
\checkmark&&& $( 8 | 5 , 6 | 2^9 )^\dagger$&$(1+4t+t^2)/{(1-t)^5}^\dagger$&\mbox{Segre $\cup$ $\IC^4$}&$LLe, LH, H\barH$\\ \hline
\checkmark&\checkmark&& {\it trivial}&{\it trivial}&\mbox{$\IC$}&$LLe$\\ \hline
\checkmark&&\checkmark& $( 8 | 5 , 6 | 2^9 )$&$(1+4t+t^2)/(1-t)^5$&\mbox{Segre} & $LLe, LH$\\ \hline
\checkmark&\checkmark&\checkmark& {\it trivial}&{\it trivial}&\mbox{$\IC$}&$LLe$\\ \hline
\end{tabular}
\end{center}}
\vspace{-0.35cm}
{\footnotesize $\dagger$ Segre branch only}
\vspace{-0.35cm}
{\caption{\label{LbarHe-table}{\bf Vacuum moduli space for superpotentials $W$ with $L\barH e$ elemental trilinear.} The description according to our standard notation, the Hilbert series, geometry, and which operators are non-vanishing in the vacuum are presented against the GIOs present in the superpotential (marked with a tick).}}
\end{table}

\subsubsection{$LLe$}

Comparing cases with superpotentials containing $LLe$ instead of $L\barH e$, we notice a couple of differences. As before, we start by considering the elemental trilinear only,
\begin{equation}
\label{WLLe}
W = C_{ijk} \, L^i_\alpha L^j_\beta \, \eps \, e^k \; ,
\end{equation}
where the coupling coefficients $C_{ijk}$ can be considered antisymmetric in the $i,j$ indices due to the $\eps$ factor. Therefore, the $e$-term equations will impose the following linear constraints on the $LL$ operators,
\begin{equation}
\left(
\begin{array}{ccc}
C_{121} & C_{131} & C_{231} \\
C_{122} & C_{132} & C_{232} \\
C_{123} & C_{133} & C_{233}
\end{array}
\right) \cdot
\left(
\begin{array}{c}
L^1_\alpha L^2_\beta \, \eps \\
L^1_\alpha L^3_\beta \, \eps \\
L^2_\alpha L^3_\beta \, \eps 
\end{array}
\right)
= 0 \; .
\end{equation}
With generic coefficients, the coefficient matrix will be non-singular. Hence, $LL$ must vanish implying that $LLe=0$ in the vacuum. This is somewhat similar to the vanishing of $L\barH$ in the previous case. However, there is a major difference intrinsic to the relations among GIOs which will lead to a different vacuum geometry.

The non-vanishing GIOs are organised in the following way:
\begin{equation}
\begin{array}{ccccccccc}
L\barH & & e && \{LH,H\barH\} & & L\barH e & & \{LH,H\barH\} \\
\IP^2 & \times & \IP^2 &\times& \IC^4&\longrightarrow & \IP^{8} &\times& \IC^4 \\
\mbox{$[x_0 : x_1 : x_2 ]$}
& &  
\mbox{$[z_0 : z_1 : z_2]$}
& &(x_3 , x_4 , x_5 , x_6)&\rightarrow & x_i z_j &&  (x_3 , x_4 , x_5 , x_6) \\
\end{array} ~,\label{LHIeLHHHI}
\end{equation}
where $i,j = 0, \ldots, 2$. The fundamental difference with the previous $L\barH e$ case is that no intrinsic relations exist between $L\barH e$ and $LH$ as was the case for $LLe$ and $LH$. The embedding~\eqref{LHIeLHHHI} simply corresponds to the variety Segre $\times$ $\IC^4$, as can be seen from direct comparison with~(\ref{Segre}).

Turning to the effect of the other F-term equations, we have from the $L^i_\alpha$-term equation,
\begin{equation}\label{Lterm}
C_{ijk} \, L^j_\beta \, \eps \, e^k  = 0\; .
\end{equation}
Since the coupling coefficients are antisymmetric in $i$ and $j$ we can redefine the electric fields, via a rotation in the field space in the following way,
\begin{equation}
\left(
\begin{array}{c}
\tilde e^1 \\
\tilde e^2 \\
\tilde e^3 \\
\end{array}
\right) =
\left(
\begin{array}{ccc}
C_{121} & C_{122} & C_{123} \\
C_{131} & C_{132} & C_{133} \\
C_{231} & C_{232} & C_{233}
\end{array}
\right) \cdot
\left(
\begin{array}{c}
e^1 \\
e^2 \\
e^3 \\
\end{array}\label{newe}
\right) \; .
\end{equation}
Thus, contracting the conditions~\eqref{Lterm} with $\barH_\beta$, we obtain
\begin{eqnarray}
\nn
L^2_\alpha \barH_\beta \, \eps \tilde e^1 +L^3_\alpha \barH_\beta \, \eps \tilde e^2 &=& 0 \\
\nn
L^1_\alpha \barH_\beta \, \eps \tilde e^1 - L^3_\alpha \barH_\beta \, \eps \tilde e^3 &=& 0 \\
- L^1_\alpha \barH_\beta \, \eps \tilde e^2 - L^2_\alpha \barH_\beta \, \eps \tilde e^3 &=& 0 \; .
\end{eqnarray}
These equations are similar to the $L\barH \tilde e$ case~\eqref{simil1}.
With an adequate choice of labeling we again have $x_i = \tilde{z}_i$ and the embedding~\eqref{LHIeLHHHI} becomes,
\begin{equation}
\begin{array}{ccccccc}
L\barH \sim \tilde e & & \{LH,H\barH\} & & L\barH \tilde e && \{LH,H\barH\}\\
\IP^2 &\times& \IC^4 & \longrightarrow & \IP^{5} &\times& \IC^4 \\
\mbox{$[x_0 : x_1 : x_2 ]$}&& (x_3 , x_4 , x_5 , x_6)
& \rightarrow & x_i x_j &&  (x_3 , x_4 , x_5 , x_6) \\
\end{array} ~,\label{LHIeLHHHIveron}
\end{equation}
where $i,j = 0, \ldots, 2$ and $i \leq j$. However, we still have an additional constraint relating $L\barH$ and $LH$. Indeed, we can contract~\eqref{Lterm} with $H_\alpha$ and multiply with $L\barH$. This leads to a set of relations involving $LH$ and $L\barH \tilde e$. In fact, we will obtain the same set of relations linking $x_0,x_1,x_2$ with $\tilde{z}_0,\tilde{z}_1,\tilde{z}_2$ for the set of variables $x_3,x_4,x_5$. Thus we have $x_{i+3}\sim x_i$ and  $x_3,x_4,x_5$ can be represented by a factor of proportionality $\lambda$. The vacuum geometry for this superpotential is thus described by,
\begin{equation}
\begin{array}{ccccccccc}
L\barH \sim \tilde e && LH && H\barH && \{L\barH \tilde e, LH\} && H\barH\\
\IP^2 & \times & \IC & \times & \IC & \longrightarrow & \IC^{9} &\times& \IC \\
\mbox{$[x_0 : x_1 : x_2 ]$} && \lambda && x_6
& \rightarrow & (x_i x_j,\lambda x_i) &&  (x_6) \\
\end{array} ~.\label{defver}
\end{equation}
The Hilbert series corresponding to the $\{L\barH\tilde e, LH\}$ part of the embedding (ignoring $x_6$) is given by,
\be\label{defVer}
\frac{(1+5t+t^2)}{(1-t)^4}.
\ee
This is a $4$-dimensional affine Calabi-Yau variety which we will refer to as the deformed Veronese (this is because the geometry reduces to the Veronese surface when $\lambda = 0$). The full geometry, including $x_6$, is of course the deformed Veronese $\times$ $\IC$.

Let us now consider deformation of the superpotential involving elemental bilinears. First, we should observe that incorporating $H\barH$ will trivialise the vacuum. Effectively, the $F_{\barH}$-term equations imply that $H=0$ and, thus, $LH=H\barH=0$ in the vacuum. From the $F_H$-term equations, we have two cases depending on whether we include $LH$ in the superpotential or not. If we do not include it, we simply have $\barH=0$ and thus $L\barH e = 0$, giving a trivial vacuum where all GIOs vanish. If we do include $LH$ in the superpotential, we obtain,
\begin{equation}
C'_iL^i_\alpha\eps + C_0 \barH_\alpha\eps =0 \; , 
\end{equation}
where $C'_i$ are the coupling constants of the $LH$ operators. Contracting this with $L^j_\beta$, and remembering that the $LL$ combinations must vanish from the $F_e$-term equations, we can conclude that $LH = 0$. Thus we again have a trivial vacuum where all GIOs vanish.

Finally, we still need to consider the last case of a superpotential with $LLe$ and $LH$,
\begin{equation}
W = C_{ijk} \, L^i_\alpha L^j_\beta \, \eps \, e^k + C'_{i} L^i_\alpha H_\beta \eps \, \; .
\end{equation}
Again, the $F_e$-term equations imply that $LLe$ must vanish. The $F_H$-term equation implies that $C'_i L^i_\alpha = 0 $. Redefining the $L$ variables by absorbing the coupling constants $C'_i$, we obtain
\begin{equation}
\label{sumL=0}
\tilde L^1_\alpha + \tilde L^2_\alpha + \tilde L^3_\alpha = 0 \; .
\end{equation}
Moreover, the $F_{L^i_\alpha}$-term equation leads to
\begin{equation}
\label{LtermH}
C_{ijk} \, L^j_\beta \, \eps \, e^k + C'_iH_\beta \eps = 0\; .
\end{equation}
Contracting with $L_\alpha^l$, and knowing that $LLe=0$, we can conclude that $LH=0$. 

Equation~(\ref{LtermH}) gives us a second relation amongst the GIOs, upon contraction with $\barH_\alpha$. For clarity of presentation, we can again absorb the coefficients $C'_i$ into the lepton doublets $\tilde{L}_{\alpha}^i$, and redefine the singlet leptons in the same manner as was done in~\eqref{newe} previously, to obtain 
\begin{eqnarray}
-\tilde L^2_\alpha \barH_\beta \, \eps \tilde e^1 -\tilde L^3_\alpha \barH_\beta \, \eps \tilde e^2 &=& H_\alpha\barH_\beta \eps \label{Lcont1} \\
\tilde L^1_\alpha \barH_\beta \, \eps \tilde e^1 - \tilde L^3_\alpha \barH_\beta \, \eps \tilde e^3 &=& H_\alpha\barH_\beta \eps \label{Lcont2}\\
\tilde L^1_\alpha \barH_\beta \, \eps \tilde e^2 + \tilde L^2_\alpha \barH_\beta \, \eps \tilde e^3 &=& H_\alpha\barH_\beta \eps \; .\label{Lcont3}
\end{eqnarray}
We thus have three equations for seven variables ($\tilde{e}$, $\tilde{L}H$, and $H\barH$). They can be interpreted as providing some linear embedding for the $H\barH$ degree of freedom, which is consequently irrelevant for the geometry. Moreover, only four degrees of freedom remains from the $L\barH e$ operators. To see this, one may take the difference of equations~(\ref{Lcont3}) and~(\ref{Lcont2}), and use the sum rule in~\eqref{sumL=0} to arrive at the constraint
\begin{equation}
\label{sume=0}
- \tilde L^1_\alpha \barH_\beta \, \eps \tilde e^1 + \tilde L^1_\alpha \barH_\beta \, \eps \tilde e^2  - \tilde L^1_\alpha \barH_\beta \, \eps \tilde e^3 = 0 \; .
\end{equation}
The other combinations arising from~(\ref{Lcont1})-(\ref{Lcont3}) simply lead to the same constraint equation with $\tilde L^1 \leftrightarrow \tilde L^2 \leftrightarrow \tilde L^3$ interchanged. In the language of the embedding~\eqref{LHIeLHHHI}, the constraints~\eqref{sumL=0} and~\eqref{sume=0} can easily be expressed in the following way. With an adequate choice of labeling $\tilde x^i=\tilde L^i_\alpha\barH_\beta\eps$ and $\tilde z_i=\pm \tilde e^i$, we have,
\begin{eqnarray}
  \nn
 \tilde z_0+ \tilde z_1+ \tilde z_2=0 \; , \\
 \tilde x_0+ \tilde x_1+ \tilde x_2=0 \; .
\end{eqnarray}
Thus, the geometry is described from the following embedding (eliminating the irrelevant $H\barH$ variables),
\begin{equation}
\begin{array}{ccccc}
\tilde L\barH & & \tilde e & & \tilde L\barH \tilde e \\
\IP^2 & \times & \IP^2 & \longrightarrow & \IP^{8} \\
\mbox{$[\tilde x_0 : \tilde x_1 : \tilde x_2 ]$}
& &  
\mbox{$[\tilde z_0 : \tilde z_1 : \tilde z_2]$}
& \rightarrow & \tilde x_i \tilde z_j \\
\end{array} ~,
\end{equation}
where the $\tilde{x}$ and $\tilde{z}$ are subject to the same relations as above.
It turns out that this geometry corresponds to a conifold, and its Hilbert series is given by
\be
\frac{(1+t)}{(1-t)^3}\, . \label{conifold}
\ee

\begin{table}[t]
{\begin{center}
\begin{tabular}{|c|c|c||c|c|c|c|}\hline
\multicolumn{3}{|c||}{$W$ \mbox{terms}} & \multicolumn{4}{c|}{\mbox{Vacuum moduli space}}\\ \hline
$LLe$ & $LH$ & $H\barH$ & \mbox{description} & \mbox{Hilbert series}  & \mbox{geometry} & \mbox{operators}
\\
\hline \hline
\checkmark&&& $( 9 | 5 , 7 | 2^{14} )$&$(1+5t+t^2)/(1-t)^5$&\mbox{def. Ver. $\times$ $\IC$} & $L\barH e, LH, H\barH$\\ \hline
\checkmark&\checkmark&& $( 3 | 3 , 2 | 2^1 )$&$(1+t)/(1-t)^3$&\mbox{Conifold} & $L\barH e, H\barH$\\ \hline
\checkmark&&\checkmark& {\it trivial}&{\it trivial}&\mbox{point at the origin}& -\\ \hline
\checkmark&\checkmark&\checkmark& {\it trivial}&{\it trivial}&\mbox{point at the origin} &-\\ \hline
\end{tabular}
\end{center}}{\caption{\label{LLe-table}{\bf Vacuum moduli space for superpotentials $W$ with $LLe$ elemental trilinear.} The description according to our standard notation, the Hilbert series, geometry, and which operators are non-vanishing in the vacuum are presented against the GIOs present in the superpotential (marked with a tick).}}
\end{table}

The vacuum geometries described in this subsection are summarised in Table~\ref{LLe-table}. Before moving on to the case where a right-handed neutrino field is included, let us compare the results in Table~\ref{LLe-table} to those of Table~\ref{LbarHe-table}. We begin to see the emergence of a clear distinction between the lepton doublet $L$ and the Higgs doublet $\barH$. Despite being embedded into the same master spaces, and despite having the same sets of non-vanishing GIOs in the vacuum, a world with $W=L\barH e$ and one with $W=LLe$ are clearly distinguished geometrically. The richer set of relations in the $LLe$ case leads to the identification of coordinates which carry unit lepton number, leading to the deformed Veronese (of affine dimension 4), as opposed to the Segre variety (of affine dimension five). In the case where $W=L\barH e$, inclusion of the bilinear $LH$ trivialised the vacuum, while inclusion of $H\barH$ has no additional effect on the vacuum. For the case $W=LLe$ the analogous circumstance does not hold completely. Here now $H\barH$ does indeed trivialise the vacuum, but inclusion of the bilinear $LH$ (with lepton number 1) is no longer ineffectual, but instead adds further linear constraints which reduces the geometry further to a conifold. Similar behavior will emerge in the case with right-handed neutrinos, though the number of possible superpotential combinations will greatly expand, making an exhaustive discussion less feasible.


\section{Right-handed Neutrinos}\label{Neut_sec}

We now consider how F-term constraints are modified when three generations of right-handed neutrinos are included in the theory. From the standpoint of the MSSM electroweak sector, a neutrino is an absolute singlet, with no charges under $SU(2)_L \times U(1)_Y$. This makes it unique among the fields we consider, in that it is itself a GIO. It also means that a so-called Majorana mass term $W \ni \nu^2$ is allowed by the gauge symmetries we are considering. 

The sterility of the neutrino under the MSSM gauge group (and its $SU(5)$ GUT extension) is simply a fact. But if the MSSM were to descend from a string construction through an intermediary stage of $SO(10)$ or $E_6$, then we would certainly {\em not} expect the neutrino to be an overall gauge singlet, and $\nu$ would not be part of the set of GIOs that establish the vacuum moduli space. Of course, in the case of $SO(10)$ there is a built in matter parity in the form of a gauged $U(1)_{B-L}$, and neutrinos are clearly identified as leptons. But in the present context the lepton number assignment of the neutrino remains ambiguous: in this case, calling $\nu$ a ``neutrino'' field presumes a Yukawa interaction of $LH\nu$ in the superpotential. In the present work we hope to understand how singlet extensions of the MSSM with a right handed-neutrino alter the geometrical classification arrived at in the previous section.

Working only at the renormalizable level, the superpotential combinations we consider can contain the standard Dirac mass term built from the $LH$ bilinear, as well as a trilinear coupling built from the $H\barH$ bilinear. That is, the superpotential couplings can now be enlarged to include the following composite trilinears:
\be
{\cal T}_\nu = \left\{H\barH \nu, LH \nu\right\} \; , \label{tricomp}
\ee
where $\nu$ are the right-handed neutrino fields. In principle we could consider superpotential terms linear, quadratic and cubic in the presumed singlet neutrino. In practice, we eliminate the tadpole-like linear term, as any such term is unlikely to arise in the superpotential from a fundamental, UV-complete theory. We will also not explicitly include the cubic $\nu^3$ term, restricting ourselves to only the putative Majorana mass term,
\be
M = \left\{\nu^2\right\} \label{nusq}
\ee
in our superpotential. More explicitly, any superpotential containing one or more operators from the set
\begin{equation} M' = \left\{\nu^2, \nu^3\right\} \end{equation}
will yield the same vacuum geometry. Thus including the $\nu^3$ as an explicit term in the superpotential is redundant.

Having established the possible GIOs and superpotential couplings in our theory, we now turn to computing the vacuum geometry for all possible choices of superpotential constructed out of these operators. It should now be clear to the reader how to derive each vacuum geometry according to the methodology from the previous examples. Therefore, for the sake of brevity, we will only present the algebraic varieties obtained without detailed analytical derivations. We begin with cases that exclude the possibility of a Majorana mass term as in~(\ref{nusq}), considering this perturbation only at the end. We then proceed by again splitting the cases according to the number of elemental trilinears present in the superpotential. 

\subsection{Cases with two elemental trilinears}

The first thing to note is that any superpotential with two elemental trilinears $LLe + L\barH e$ again lead to trivial vacuum geometries, in the sense that it is composed of points and/or lines. Again, the corresponding GIOs $LLe$ and $L\barH e$ are lifted from the vacuum and the remaining combinations of $LH, H\barH$ and $\nu$ are only present as invariant polynomials that do not have any relations and syzygies among themselves. We thus arrive at a situation analogous to Table~\ref{two-trilin} with the vacuum moduli space of $\mathbb{C}^7$ for the case without bilinear deformation. Addition of a $LH$ or $H\barH$ removes these bilinears, leaving $\mathbb{C}^3$, parameterized by the undetermined neutrino fields.

\subsection{Cases without any elemental trilinear}

The vacuum moduli space for all possible cases without any elemental trilinear are presented in Table~\ref{nu-no-tril}. Due to the composite nature of some of the superpotential terms, as in~(\ref{tricomp}), some vacua can be constituted of two non-trivial branches. We can also observe that all branches have palindromic Hilbert series and are thus Calabi-Yau varieties. This was also true before we introduced right-handed neutrinos, as is evident from Table~\ref{no-trilin}. For the sake of brevity, from this point onward we will only present the numerator coefficients of the Hilbert series, ordered according the degree of $t$. That is, $(a,b,c,\dots)$ means $a+bt+ct^2+\ldots$, and the numerator always takes the form $(1-t)^d$ where $d$ is the dimension of the variety.

\begin{table}[t]
{\begin{center}
\begin{tabular}{|c||c|c|c|c|c|}\hline
\multicolumn{1}{|c||}{$W$ \mbox{terms}} & \multicolumn{5}{c|}{\mbox{Vacuum moduli space}}\\ \hline
 & \multicolumn{2}{|c|}{first branch} & \multicolumn{2}{c|}{second branch} & \multirow{2}{*}{\mbox{operators}} \\ \cline{2-5}
 & \mbox{HS} & \mbox{dim} & \mbox{HS} & \mbox{dim}& \\ \hline \hline
$LH \nu$ & $(1,13,28,13,1)$&$8$&$(1,4,1)$ &$6$& $LLe, L\barH e, H\barH, \nu$\\ \hline
$LH (\nu+1)$ & $(1,13,28,13,1)$&$8$&$(1,4,1)$ &$6$& $LLe, L\barH e, H\barH, \nu$\\ \hline
$LH \nu+H\barH$ & $(1,4,1)$  &$8$& - & - &$LLe, L\barH e, \nu$\\ \hline
$LH (\nu+1)+H\barH$ &$(1,4,1)$ &$8$& - & - & $LLe, L\barH e, \nu$\\ \hline
$H\barH \nu$ &  $(1,13,28,13,1)$&$10$&$(1,4,1)$ &$8$& $LLe, L\barH e, LH, \nu$\\ \hline
$H\barH (\nu+1)$ &  $(1,13,28,13,1)$&$10$&$(1,4,1)$ &$8$& $LLe, L\barH e, LH, \nu$\\ \hline
$H\barH \nu+LH$ &$(1,4,1)$ &$8$& - & - & $LLe, L\barH e, \nu$\\ \hline
$H\barH (\nu+1)+LH$ &$(1,4,1)$ &$8$& - & - & $LLe, L\barH e, \nu$\\ \hline
$(H\barH + LH)\nu$ &  $(1,13,28,13,1)$&$8$&$(1,4,1)$ &$6$& $LLe, L\barH e, H\barH, \nu$\\ \hline
$H\barH \nu + LH(\nu+1)$ & $(1,13,28,13,1)$ &$8$& - & - & $LLe, L\barH e, \nu$\\ \hline
$H\barH (\nu+1) + LH\nu$ & $(1,13,28,13,1)$ &$8$& - & - & $LLe, L\barH e, \nu$\\ \hline
$(H\barH + LH)(\nu+1)$ & $(1,13,28,13,1)$ &$8$& - & - & $LLe, L\barH e, \nu$\\ \hline
\end{tabular}
\end{center}}{\caption{\label{nu-no-tril}{\bf Vacuum moduli space for superpotentials without any elemental trilinears, including possible bilinear and composite trilinear deformations}. The Hilbert series (HS) and dimension (dim) of the different branches of the algebraic varieties are presented in relation to the operators present in the superpotential. The final column lists the type of GIOs that do not vanish in the vacuum. Each of the $(1,13,28,13,1)$ varieties are made out of $63$ quadratics and each of the $(1,4,1)$ varieties are made out of $9$ quadratics. We would also remind the reader that the degree is easily obtained from the sum of the HS coefficients. Hence the $(1,4,1)$ are of degree $6$ and the other ones are of degree $56$.}}
\end{table}

In the present case, only two geometries occur. First, we find the Segre variety again with the Hilbert series coefficients $(1,4,1)$ in the numerator. These are the cases from Table~\ref{no-trilin}. The variation in dimension is simply given by $\mbox{Segre} \times \IC^n$, where $n=1,3$ to obtain the corresponding dimensions. These additional flat directions are given by the three neutrino fields $\nu$ or the $H\barH$ operator, for which no relations are possible with $LLe$ or $L\barH e$.

The other branches are given by the Hilbert series $(1,13,28,13,1)$ which is the same as the ideal of relations among the GIOs $\left\{LLe, L\barH e, LH, H\barH\right\}$. However, in the neutrino case, we have an eight-dimensional variety constituted of $\left\{LLe, L\barH e, \nu\right\}$. We also have a $10$-dimensional one which can be decomposed as ${\rm CY}_8 \times \IC^2$, where ${\rm CY}_8$ is an eight-dimensional Calabi-Yau with Hilbert series $(1,13,28,13,1)$ built out of $\left\{LLe, L\barH e, LH\right\}$. The extra $\IC^2$ comes from neutrino flat directions.

The appearance of this eight-dimensional Calabi-Yau ${\rm CY}_8$ is remarkable, in so far as the Hilbert series numerator $(1,13,28,13,1)$ is also that of the nine-dimensional Calabi-Yau ${\rm CY}_9$ that defined the vacuum moduli space of the MSSM electroweak sector~(\ref{BigCYHS}). Recall that this was the vacuum manifold defined by relations amongst the GIOs in the set ${\cal S}'' = \{LLe, L\barH e, LH, H\barH\}$, when $W=0$. The defining polynomials of the ${\rm CY}_8$ are those of the ${\rm CY}_9$ intersected with the surface defined by $H\barH = 0$. But, for the cases with no $LH$ in the vacuum, the role of the $LH$ GIOs in defining the vacuum manifold is being taken by the neutrino GIOs.



In comparing the various entries in Table~\ref{nu-no-tril}, it is clear that in the absence of fundamental trilinears in the superpotential, there is no distinction between $L$ and $\barH$ in terms of vacuum geometry. This is akin to the perfect symmetry in assigning lepton number to the $SU(2)$ doublets that obtained in Sections~\ref{sec:31} and~\ref{sec:32}. Another way of stating this finding is that there is nothing geometrically special about the standard Dirac mass term for the neutrino $W \ni L H \nu$ in the absence of other fundamental trilinears in the superpotential. Their presence will be the subject of the next subsection.

Finally, we note that the vacuum geometries in Table~\ref{no-trilin} come in pairs. That is, when a composite trilinear is present in the superpotential, the addition of the corresponding fundamental bilinear has no effect on the overall geometry. So, for example, $W=LH\nu$ gives rise to the same geometry as $W=LH + LH\nu$, which we denote in the table by the shorthand notation $LH(\nu+1)$. This again gives support to the notion that the neutrino Dirac operator carries no particular geometrical significance at this stage in our analysis. This property will continue to hold, even in the presence of fundamental trilinears in the superpotential, but will be altered when Majorana mass terms for the neutrino are introduced.

\subsection{Cases with only one elemental trilinear}

\subsubsection{$L\barH e$}

Let us consider cases with the $L\barH e$ operators first. Results are presented in Table~\ref{nu-LHe}. Again, the $F_e$-term equations lift the $L\barH e$ operators from the vacuum in all cases. As with Table~\ref{LbarHe-table}, inclusion of the fundamental bilinear $LH$, alone, will continue to produce a trivial background. But by including the composite trilinear $LH\nu$ the geometry is altered by relations imposed by the new $F_{\nu}$-term equations. All superpotentials in Table~\ref{nu-LHe} which contain this Dirac operator for the neutrino result in a deformed Veronese~\eqref{defVer}, independent of other operators that may be present (fundamental bilinears or composite trilinears). But the realization of the deformed Veronese variety in terms of the underlying GIOs is markedly different than in Section~\ref{secLHe}. In this case, the field $\nu$ takes the role of $LH$ in the embedding, in the same sense that we saw in the previous subsection. Specifically, the embedding is given by:
\begin{equation}
\begin{array}{cccccc}
LL \sim \tilde e && \nu &&& \{LL \tilde e, \nu\}\\
\IP^2 & \times & \IC && \longrightarrow & \IC^{9}\\
\mbox{$[x_0 : x_1 : x_2 ]$} && \lambda && \rightarrow & (x_i x_j,\lambda x_i)\\
\end{array} ~,\label{defvernu}
\end{equation}
with $i,j=1,2,3$. See~\cite{He:2014oha} for details.

\begin{table}[t]
{\begin{center}
\begin{tabular}{|c||c|c|c|c|c|}\hline
\multicolumn{1}{|c||}{$W$ \mbox{terms}} & \multicolumn{5}{c|}{\mbox{Vacuum moduli space}}\\ \hline
\multirow{2}{*}{$\bm{L\barH e \, + }$} & \multicolumn{3}{|c|}{top component} & \multicolumn{2}{c|}{full space}   \\ \cline{2-6}
 & \mbox{description} & \mbox{HS} & \mbox{dim} & \mbox{geometry}& \mbox{operators}
\\
\hline \hline
$H\barH \nu$ & $( 11 | 8 , 14 | 3^{6}2^3 )$&$(1,4,7,2)$&$8$&\mbox{non-Calabi-Yau}&$LLe, LH, \nu$\\ \hline
$H\barH (\nu+1)$ & $( 11 | 8 , 14 | 3^{6}2^3 )$&$(1,4,7,2)$&$8$&\mbox{non-Calabi-Yau}&$LLe, LH, \nu$\\ \hline
$H\barH \nu+LH$ & {\it trivial} & {\it trivial} &$4$&$\IC^4$&$LLe, \nu$\\ \hline
$H\barH (\nu+1)+LH$ & {\it trivial} & {\it trivial} &$4$&$\IC^4$&$LLe, \nu$\\ \hline
$LH \nu$ & $( 8 | 4 , 7 | 2^{14} )$&$(1,5,1)$&$4$&${\rm def.~Ver.}\cup\IC $&$LLe, H\barH, \nu$\\ \hline
$LH (\nu+1)$ & $( 8 | 4 , 7 | 2^{14} )$&$(1,5,1)$ &$4$&${\rm def.~Ver.}\cup\IC $&$LLe, H\barH, \nu$\\ \hline
$LH \nu+H\barH$ & $( 8 | 4 , 7 | 2^{14} )$&$(1,5,1)$ &$4$&${\rm def.~Ver.}$&$LLe, \nu$\\ \hline
$LH (\nu+1)+H\barH$ & $( 8 | 4 , 7 | 2^{14} )$&$(1,5,1)$ &$4$&${\rm def.~Ver.}$&$LLe, \nu$\\ \hline
$(H\barH +LH) \nu$ & $( 8 | 4 , 7 | 2^{14} )$&$(1,5,1)$&$4$&${\rm def.~Ver.}\cup\IC $&$LLe, LH, H\barH, \nu$\\ \hline
$H\barH \nu+LH (\nu+1)$ & $( 8 | 4 , 7 | 2^{14} )$&$(1,5,1)$&$4$&${\rm def.~Ver.}$&$LLe, \nu$\\ \hline
$H\barH (\nu+1)+LH \nu$ & $( 8 | 4 , 7 | 2^{14} )$&$(1,5,1)$&$4$&${\rm def.~Ver.}$&$LLe, \nu$\\ \hline
$(H\barH+LH)(\nu+1)$ & $( 8 | 4 , 7 | 2^{14} )$&$(1,5,1)$&$4$&${\rm def.~Ver.}$&$LLe, \nu$\\ \hline
\end{tabular}
\end{center}}{\caption{\label{nu-LHe}{\bf Vacuum moduli space for superpotentials with $L\barH e$ elemental trilinear, including possible bilinear and composite trilinear deformations.} The description, Hilbert series (HS) and dimension (dim) of the top component are presented. The geometry and the type of operators that do not vanish in the vacuum corresponds to the full moduli space.}}
\end{table}

A new vacuum geometry also appears for the first time in Table~\ref{nu-LHe}, for the cases where the $LH$ operator is absent from the superpotential (as a stand-alone blinear or as part of a composite trilinear). The variety is an affine cone over a degree-14 projective variety defined by the intersection of three quadratics with six cubic polynomials. This is the only instance in our study in which the variety is defined by a set of cubic relations.
The corresponding Hilbert series is
\begin{equation}
\frac{(1+4t+7t^2+2t^3)}{(1-t)^8}\, . \label{1472}
\end{equation}
The variety is not Calabi-Yau since the Hilbert series is not palindromic.

\subsubsection{$LLe$}

Let us now turn to the cases involving the $LLe$ operator. All results are presented in Table~\ref{LLe}. The $LLe$ GIOs are lifted from the vacuum and we again find a trivial vacuum when including both $LLe$ and $H\barH$ terms, provided that no $H\barH\nu$ terms are present, in complete analogy with the previous case with $L\barH e$ and $LH$. We also have the same non-Calabi-Yau variety with Hilbert series coefficients $(1,4,7,2)$ for the case with $LH\nu$ and no $H\barH$ operators at all, very similarly to the previous case. However, the dimension of this space is now $6$, as opposed to $8$. This can be accounted by the fact that the $LH$ operators are being replaced by $H\barH$ and thus two less degrees of freedom are present.

\begin{table}[t]
{\begin{center}
\begin{tabular}{|c||c|c|c|c|c|}\hline
\multicolumn{1}{|c||}{$W$ \mbox{terms}} & \multicolumn{5}{c|}{\mbox{Vacuum moduli space}}\\ \hline
\multirow{2}{*}{$\bm{LLe \, + }$} & \multicolumn{3}{|c|}{top component} & \multicolumn{2}{c|}{full space}   \\ \cline{2-6}
 & \mbox{description} & \mbox{HS} & \mbox{dim} & \mbox{geometry}& \mbox{operators}
\\
\hline \hline
$LH \nu$ & $( 9 | 6 , 14 | 3^{6}2^3 )$&$(1,4,7,2) $&$6$&\mbox{non-Calabi-Yau}&$L\barH e, H\barH, \nu$\\ \hline
$LH (\nu+1)$ & $( 9 | 6 , 14 | 3^{6}2^3 )$&$(1,4,7,2) $&$6$&\mbox{non-Calabi-Yau}&$L\barH e, H\barH, \nu$\\ \hline
$LH \nu+H\barH$ & {\it trivial} & {\it trivial} &$3$&$\IC^3$&$\nu$\\ \hline
$LH (\nu+1)+H\barH$ & {\it trivial} & {\it trivial} &$3$&$\IC^3$&$\nu$\\ \hline
$H\barH \nu$ & $( 10 | 6 , 7 | 2^{14} )$&$(1,5,1)$&$6$&${\rm def.~Ver.}\times \IC^2\cup\IC^3$&$L\barH e,LH,\nu$\\ \hline
$H\barH (\nu+1)$ & $( 10 | 6 , 7 | 2^{14} )$&$(1,5,1)$&$6$&${\rm def.~Ver.}\times \IC^2\cup\IC^3$&$L\barH e,LH,\nu$\\ \hline
$H\barH \nu+LH$ & $( 4 | 4 , 2 | 2^1 )$&$(1,1)$&$4$&${\rm Conifold}^-\times\IC^2\cup\IC^3$&$L\barH e, \nu$\\ \hline
$H\barH (\nu+1)+LH$ & $( 4 | 4 , 2 | 2^1 )$&$(1,1)$&$4$&${\rm Conifold}^-\times\IC^2\cup\IC^3$&$L\barH e, \nu$\\ \hline
$(H\barH +LH)\nu$&$(7|4,6|2^{9})$&$(1,4,1)$ &$4$&$\mbox{Segre}^-\cup\IC^3\cup\IC^2$&$L\barH e,LH,H\barH,\nu$\\ \hline
$H\barH \nu+LH (\nu+1)$ & $( 7 | 4 , 6 | 2^{9} )$&$(1,4,1)$ &$4$&$\mbox{Segre}^-\cup\IC^3$&$L\barH e, \nu$\\ \hline
$H\barH (\nu+1)+LH \nu$ & $( 7 | 4 , 6 | 2^{9} )$&$(1,4,1)$ &$4$&$\mbox{Segre}^-\cup\IC^3$&$L\barH e, \nu$\\ \hline
$(H\barH+LH)(\nu+1)$ & $( 7 | 4 , 6 | 2^{9} )$&$(1,4,1)$ &$4$&$\mbox{Segre}^-\cup\IC^3$&$L\barH e, \nu$\\ \hline
\end{tabular}
\end{center}}{\caption{\label{LLe}{\bf Vacuum moduli space for superpotentials with $LL e$ elemental trilinear, including possible bilinear and composite trilinear deformations.} The description, Hilbert series (HS) and dimension (dim) of the top component are presented. The geometry and the type of operators that do not vanish in the vacuum corresponds to the full moduli space.
The $^-$ superscript denotes spaces analogous to the one indicated, but of one dimension less.
}}
\end{table}

In fact, the first six entries in Table~\ref{LLe} share the same vacuum variety as the first six entries in Table~\ref{nu-LHe}, which can be obtained via a swap of fields with opposite R-parity assignment, $\barH \leftrightarrow L$. However, the remaining cases in Table~\ref{LLe} present differences with the $L\barH e$ cases of the previous section. We find no further instances of the deformed Veronese surface, but instead obtain new vacuum structures. The defining polynomials and Hilbert series data are those of the conifold and the Segre embedding. However, the dimensions are always one less than the usual varieties, despite being described by similar polynomials. That is, we find the Hilbert series~(\ref{conifold}) and~(\ref{H_EW}), but with denominators $(1-t)^2$ and $(1-t)^4$, respectively. Therefore, we denote these spaces with the superscript $^-$ to differentiate them from the Segre and conifold.

\subsection{Majorana mass term}

Thus far, the singlet object denoted by ``$\nu$'' shows no particular geometric predilection for behaving like the traditional neutrino. The coupling represented by the Yukawa interaction $W\ni LH\nu$ does not prefer a traditional lepton number assignment for $\nu$. In the absence of fundamental trilinears in the superpotential, both putative lepton number assignments (equivalently, both possible R-parity assignments) yield identical results. This was the principal conclusion drawn from Table~\ref{nu-no-tril}. There is a distinction between, for example, $W= L\barH e + LH\nu$ (which gives a deformed Veronese), and $W=LLe + LH\nu$ (which gives the six-dimensional non-Calabi-Yau geometry). But overall there seems to be nothing about Tables~\ref{nu-LHe} and~\ref{LLe} that suggests a clear lepton number assignment for $\nu$.

\begin{table}[th]
{\begin{center}
\begin{tabular}{|c||c|c|c|c|c|}\hline
\multicolumn{1}{|c||}{$W$ \mbox{terms}} & \multicolumn{5}{c|}{\mbox{Vacuum moduli space}}\\ \hline
\multirow{2}{*}{$\bm{\nu\nu \, + }$} & \multicolumn{2}{|c|}{first branch} & \multicolumn{2}{c|}{second branch} & \multirow{2}{*}{\mbox{operators}} \\ \cline{2-5}
 & \mbox{HS} & \mbox{dim} & \mbox{HS} & \mbox{dim}& \\ \hline \hline
$LH \nu$ &  $(1,11,15,3)$&$7$&$(1,4,1)$&$6$&  $ LLe, L\barH e, H\barH$\\ \hline
$LH (\nu+1)$ & $(1,4,1)$&$5$&$(1,4,1)$&$6$& $LLe, L\barH e, H\barH, \nu$\\ \hline
$LH \nu+H\barH$ &$(1,4,1)$&$5$& - & - &$LLe$\\ \hline
$LH (\nu+1)+H\barH$ &$(1,4,1)$ &$5$& - & - &$LLe, L\barH e$\\ \hline
$H\barH \nu$ & $(1,13,28,13,1)$&$8$& - & - &$LLe, L\barH e, LH$\\ \hline
$H\barH (\nu+1)$ & $(1,13,28,13,1)$&$8$&$(1,4,1)$&$5$&$LLe, L\barH e, LH, \nu$\\ \hline
$H\barH \nu+LH$ &$(1,4,1)$&$5$& - & - &$ LLe, L\barH e$\\ \hline
$H\barH (\nu+1)+LH$ &$(1,4,1)$&$5$& - & - &$LLe, L\barH e$\\ \hline
$(H\barH+LH)\nu$ &  $(1,11,15,3)$&$7$&$(1,4,1)$ &$6$&$LLe, L\barH e, H\barH$ \\ \hline
$H\barH \nu + LH(\nu+1)$ & $(1,4,1)$&$5$& - & - & $LLe, L\barH e$\\ \hline
$H\barH(\nu+1) + LH\nu$ & $(1,4,1)$&$5$& - & - &$LLe$\\ \hline
$(H\barH + LH)(\nu+1)$ &$(1,4,1)$&$5$& - & - &$ LLe, L\barH e$\\ \hline
\end{tabular}
\end{center}}{\caption{\label{nusq-notri}{\bf Vacuum moduli space for superpotentials including Majorana mass term, without any elemental trilinears.} The Hilbert series (HS) and dimension (dim) of the different branches of the algebraic varieties are presented in relation to the GIO types present in the superpotential. The operators correspond to the type of operators that do not vanish in the vacuum. Each $(1,13,28,13,1)$ variety is of degree $56$ and made out of $63$ quadratics. Each $(1,4,1)$ variety is of degree $6$ and made out of $9$ quadratics. Each $(1,11,15,3)$ variety is of degree $30$ and made out of $51$ quadratics.}}
\end{table}

The addition of a quadratic term $W\ni \nu^2$, or Majorana mass term, for the neutrino candidate sharpens the issue considerably. Such a term is commonly invoked to incorporate the see-saw mechanism for neutrino masses, though it is not strictly necessary to explain the neutrino sector of the Standard Model. From the physics point of view, if R-parity is nothing more than a discrete symmetry, here effectively amounting to the rule that all allowed operators must have an even number of ``leptons'', then such an operator is consistent with the traditional R-parity assignment when the field $\nu$ is designated as a lepton.
From the geometrical point of view, including such a Majorana term has the effect of lifting, either partially or completely, the neutrino degrees of freedom from the vacuum manifold. As a result, some of the vacuum structures in Tables~\ref{nu-no-tril} through~\ref{LLe} will change. We will here ask whether these changes tend to favor the cases in which the superpotential is consistent with the traditional R-parity assignments. Our results are presented in Table~\ref{nusq-notri} (no fundamental trilinears) and Table~\ref{nusq-tri} (cases involving $L\barH e$ or $LLe$).

Two kinds of geometries appear which where not present before. In Table~\ref{nusq-notri}, which gives results for superpotentials containing no elemental trilinears, we find the $7$-dimensional variety constituted of the relations and syzygies among $\{LLe, L\barH e\}$ operators, as presented in~\eqref{Mattifold_4}. This is the case for the two superpotentials $W=\nu\nu+LH\nu$ and $W=\nu\nu+LH\nu+H\barH\nu$. The remaining vacuum varieties for the case without elemental trilinears are ${\rm CY}_8$, Segre and $\mbox{Segre}\times \IC$.

The other geometry that appears for the first time is the Veronese variety in Table~\ref{nusq-tri}. This solution was identified in the original paper in this series~\cite{Gray:2005sr}, and was exhaustively studied much more recently~\cite{He:2014loa}. Here we find six instances of this vacuum moduli space: four of them are obtained with the $L\barH e$ elemental trilinear and the remaining two with the $LLe$ trilinear. We note that the second instance of the Veronese surface -- the third line in Table~\ref{nusq-tri} -- is the standard superpotential of the MSSM electroweak theory with a Majorana mass term.

\begin{table}[p]
{\begin{center}
\begin{tabular}{|c||c|c|c|c|c|}\hline
\multicolumn{1}{|c||}{$W$ \mbox{terms}} & \multicolumn{5}{c|}{\mbox{Vacuum moduli space}}\\ \hline
\multicolumn{6}{c}{}\\\hline
\multirow{2}{*}{$\bm{L\barH e +\nu\nu \, + }$} & \multicolumn{3}{|c|}{top component} & \multicolumn{2}{c|}{full space}   \\ \cline{2-6}
 & \mbox{description} & \mbox{HS} & \mbox{dim} & \mbox{geometry}& \mbox{operators}
\\
\hline \hline
$LH \nu$ & $( 5 | 3 , 4 | 2^{6} )$&$(1,3) $ &$3$&\mbox{Veronese  $\cup$ $\IC $}&$LLe,LH,H\barH,\nu$\\ \hline
$LH (\nu+1)$ & $( 2 | 2 , 2 | 2^{1} )$&$(1,1) $&$2$ &$\mbox{Conifold}^-\cup\IC\cup\IC$&$LLe,LH,H\barH,\nu$\\ \hline
$LH \nu+H\barH$ & $( 5 | 3 , 4 | 2^{6} )$&$(1,3) $&$3$ &\mbox{Veronese }&$LLe,LH,\nu$\\ \hline
$LH (\nu+1)+H\barH$ & $( 2 | 2 , 2 | 2^{1} )$&$(1,1) $ &$2$&$\mbox{Conifold}^-\cup\IC$&$LLe,LH,\nu$\\ \hline
$H\barH \nu$ & $( 8 | 5 , 6 | 2^9 )$&$(1,4,1)$&$5$&\mbox{Segre}&$LLe,LH,H\barH,\nu$\\ \hline
$H\barH (\nu+1)$ & $( 8 | 5 , 6 | 2^9 )$&$(1,4,1)$&$5$&\mbox{Segre $\cup$ $\IC^3 $}&$LLe,LH,H\barH,\nu$\\ \hline
$H\barH \nu+LH$ & {\it trivial} & {\it trivial} &$1$&$\IC$&$LLe$\\ \hline
$H\barH (\nu+1)+LH$ & {\it trivial} & {\it trivial} &$1$&$\IC$&$LLe$\\ \hline
$(H\barH +LH) \nu$ & $( 5 | 3 , 4 | 2^{6} )$&$(1,3) $&$3$&\mbox{Veronese $\cup$ $\IC $}&$LLe,LH,H\barH,\nu$\\ \hline
$H\barH \nu+LH (\nu+1)$ & $( 2 | 2 , 2 | 2^{1} )$&$(1,1) $&$2$&$\mbox{Conifold}^-\cup\IC$&$LLe,LH,\nu$\\ \hline
$H\barH (\nu+1)+LH \nu$ & $( 5 | 3 , 4 | 2^{6} )$&$(1,3) $&$3$&\mbox{Veronese}&$LLe,LH,\nu$\\ \hline
$(H\barH +LH)( \nu+1)$ & $( 2 | 2 , 2 | 2^{1} )$&$(1,1) $&$2$&$\mbox{Conifold}^-\cup\IC$&$LLe,LH,\nu$\\ \hline
\multicolumn{6}{c}{}\\\hline
\multirow{2}{*}{$\bm{LLe +\nu\nu \, + }$} & \multicolumn{3}{|c|}{top component} & \multicolumn{2}{c|}{full space}   \\ \cline{2-6}
 & \mbox{description} & \mbox{HS} & \mbox{dim} & \mbox{geometry}& \mbox{operators}
\\
\hline \hline
$LH \nu$ & $( 6 | 4 , 4 | 2^{6} )$&$(1,3)$&$4$&$\mbox{Veronese}\times\IC$&$L\barH e, H\barH$\\ \hline
$LH (\nu+1)$ & $( 3 | 3 , 2 | 2^1 )$&$(1,1)$&$3$&$\mbox{Conifold}^-\times \IC\cup\IC^2 $&$L\barH e,H\barH,\nu$\\ \hline
$LH \nu+H\barH$ & {\it trivial} & {\it trivial} &$0$&\mbox{point}&-\\ \hline
$LH (\nu+1)+H\barH$ & {\it trivial} & {\it trivial} &$0$&\mbox{point}&-\\ \hline
$H\barH \nu$ & $( 8 | 4 , 7 | 2^{14} )$&$(1,5,1)$&$4$&\mbox{def. Ver.}&$LH, L\barH e$\\ \hline
$H\barH(\nu+1)$ & $(8|4,7|2^{14})$&$(1,5,1)$&$4$&$\mbox{def. Ver.}\cup\mbox{point}$&$LH, L\barH e, \nu$\\ \hline
$H\barH \nu+LH$ & $( 2 | 2 , 2 | 2^1 )$&$(1,1)$&$2$&$\mbox{Conifold}^-$&$L\barH e$\\ \hline
$H\barH(\nu+1)+LH$ & $( 2 | 2 , 2 | 2^1 )$&$(1,1)$&$2$&$\mbox{Conifold}^-\cup \mbox{point}$&$L\barH e,\nu$\\ \hline
$(H\barH +LH) \nu$ & $( 5 | 3 , 4 | 2^{6} )$&$(1,3)$ &$3$&\mbox{Veronese  $\cup$ $\IC^2  $}&$L\barH e, H\barH$\\ \hline
$H\barH \nu+LH (\nu+1)$ & $( 2 | 2 , 2 | 2^{1} )$&$(1,1)$ &$2$&$\mbox{Conifold}^-$&$L\barH e$\\ \hline
$H\barH(\nu+1)+LH\nu$ & $(2|2,2|2^{1} )$&$(1,1)$&$2$&$\mbox{Conifold}^-\cup\mbox{point}$&$L\barH e,\nu$\\ \hline
$(H\barH+LH)(\nu+1)$ & $(2|2,2|2^{1} )$&$(1,1)$ &$2$&$\mbox{Conifold}^-\cup\mbox{point}$&$L\barH e,\nu$\\ \hline
\end{tabular}
\end{center}}{\caption{\label{nusq-tri}{\bf Vacuum moduli space for superpotentials with Majorana mass term and one elemental trilinear.} The description, Hilbert series (HS) and dimension (dim) of the top component are presented. The geometry and the type of operators that do not vanish in the vacuum corresponds to the full moduli space.}}
\end{table}

The addition of a Majorana operator $\nu^2$ does not trivialize any geometries from Tables~\ref{nu-no-tril}, \ref{nu-LHe}~and~\ref{LLe} that were previously non-trivial. In the absence of fundamental trilinears (Table~\ref{nu-no-tril} and~\ref{nusq-notri}), the impact of adding the Majorana mass term is rather muted, but (signficantly), its impact differentiates between superpotentials involving the lepton doublet and the Higgs doublet, thus breaking the perfect symmetry between these fields seen in Table~\ref{nu-no-tril}. 

For example, $W=LH\nu$ and $W=H\barH\nu$ both led to the vacuum moduli space CY$_8$ $\times$ Segre, in the absence of the Majorana term. Its inclusion deforms the case $W=LH\nu$ to the non-Calabi-Yau variety defined by the Hilbert series~(\ref{LLeLHIe}) and embedding~(\ref{Mattifold_4}), but merely eliminates the Segre component for the case $W=H\barH\nu$. The case $W=LH(\nu+1)$ is deformed even further, from CY$_8$ $\times$ Segre to Segre $\times$ Segre, while only the dimension of the second Segre factor changes for $W=H\barH(\nu+1)$.
Thus, in this simple system of superpotentials, the inclusion of a Majorana mass term can distinguish between pairs of cases in which $L \leftrightarrow \barH$, but not every such pair sees a distinction, and the differences are subtle.

We next compare the results involving elementral trilinears: Tables~\ref{nu-LHe} and~\ref{LLe} with the collected results in Table~\ref{nusq-tri}. In the previous subsection we observed that non-trivial entries in Table~\ref{nu-LHe} and Table~\ref{LLe} which are related by $L \leftrightarrow \barH$ interchange do show differences in the vacuum geometry, at least in some instances. The inclusion of the Majorana mass term makes these differences more stark. It is intriguing that the unique non-Calabi-Yau variety arising from cubic relations, with Hilbert series~(\ref{1472}), is deformed to the previously analyzed Segre variety for $W= L\barH e + HH\nu +\nu^2$ and $W= L\barH e + HH(\nu+1)+\nu^2$, to the Veronese surface for $W= LL e + LH\nu+\nu^2$, and to the two-dimensional conifold$^-$ variety for $W= LL e + LH(\nu+1)+\nu^2$. Without the Majorana mass term, all of these superpotentials produced the outcome defined by Hilbert series~(\ref{1472}). 

So the addition of the Majorana mass term can produce a distinction between a superpotential $W$ and its `R-parity dual' $\widetilde{W}(L \leftrightarrow \barH)$ when none existed in its absence. It can also make pre-existing distinctions more complex. Let us consider the fate of entries~5-8 in Table~\ref{nu-LHe}. In all of these cases (which include the MSSM EW sector with Dirac neutrino masses) the vacuum geometry is that of the deformed Veronese. All four entries change when the Majorana mass term is included. The traditional MSSM-like cases, $W= L\barH e + LH\nu + \nu^2$ and $W= L\barH e + LH\nu + \nu^2 + H\barH$, each become the Veronese surface, while the addition of the R-parity violating term $LH$ reduces the variety to the conifold$^{-}$.
The R-parity dual cases $\widetilde{W}$ are the entries~5-8 in Table~\ref{LLe}. The addition of the Majorana mass term has far less impact on these cases, and the Veronese surface never appears in this subset when the elemental trilinear is $LLe$.

This subset of cases is particularly relevant in that it contains the superpotential for the MSSM electroweak sector with a see-saw mechanism for the right-handed neutrinos. Both the case with and without a bilinear for the Higgs fields produce the Veronese surface, whereas inclusion of R-parity violating terms such as $LLe$ and $LH$ would produce a trivial vacuum or a 2D conifold, respectively. Only at this stage in our analysis does some geometrical preference start to emerge for the conventional R-parity-conserving MSSM electroweak sector -- the Veronese variety prefers the fundamental trilinear $L\barH e$, though the presence of a fundamental Higgs bilinear is not particularly preferred, with its the cases $H\barH \nu$ and $H\barH (\nu+1)$ also producing the Veronese variety.
These latter cases leave the lepton-number assignment of the neutrino ambiguous, but continue to forbid fundamental operators such as $LH$ and $LLe$ in the superpotential. 



\section{Discussion and outlook}\label{Discussion_sec}


This paper provides a complete classification of the vacuum geometries of all possible renormalizable theories built from the fields of the MSSM electroweak sector. This represents the culmination of a research program that began one decade ago~\cite{Gray:2005sr,Gray:2006jb}. At that time, computational algebraic geometry techniques allowed for a determination of the dimension of the vacuum moduli space, and little else. Today a much richer understanding of these spaces is possible. The non-trivial geometries in a system as simple as the $SU(2)_L\times U(1)_Y$ sector of the MSSM are surprisingly diverse, comprising Severi varieties, Calabi-Yau spaces, conifolds of various dimension, and various deformations on these spaces. 

For the most part, the geometries reflect the basic properties assumed in establishing the relevant gauge invariant operators for describing the vacuum space. That is, we here assume nothing more than what is necessary within the MSSM itself. Therefore, when discussing right-handed neutrino fields we treat them as true singlet states, as would be expected within an $SU(5)$ grand unified model. To establish the allowed couplings of this singlet field, it is helpful to consider what couplings are allowed under the premise of a conserved parity assignment (matter parity) which is closely related to lepton number. In the context of the full MSSM this would be promoted to the discrete symmetry known as R-parity.

The complete classification achieved in this work includes cases for which the standard R-parity is conserved, and those for which some R-parity violation is allowed. In a related vein, it also includes cases for which an unambiguous (conserved) lepton number can be assigned to the right handed neutrino, and cases in which the superpotential would make such an assignment ambiguous, or lead to the non-conservation of this lepton number. We find non-trivial geometries emerge in all of these superpotential categories.

Distinctions between a superpotential $W$ and its R-parity dual $\widetilde{W}$, obtained via the transposition $L \leftrightarrow \barH$, do indeed emerge, even in the absence of right-handed neutrinos, in certain circumstances. We began by studying the vacuum geometries determined by the relations that arise solely among the GIOs of the theory. Intriguingly, the geometries generated by the blilinear portion of the GIOs show precisely zero distinction between the fields $L$ and $\barH$. The algebraic variety defined by the relations amongst the $LLe$ GIOs and the $L\barH e$ GIOs are also identical, but in this case this is an accidental symmetry, made possible by the simple fact that there are three species of $L$ fields, but a single $\barH$ field in the MSSM.

When we add F-term constraints, we continue to see that no distinction between $L$ and $\barH$ emerges when the superpotential contains only fundamental, gauge-invariant bilinear terms. Allowing for composite trilinears, in which the singlet field $\nu$ is allowed to couple to fundamental bilinears, producing terms like $LH\nu$ and $H\barH \nu$, continues to exhibit a fundamental invariance under the R-parity swap $L \leftrightarrow \barH$. If the superpotential is further augmented to include a Majorana mass term , then these cases do produce slightly different vacuum geometries, but it would be difficult to argue that any one particular configuration was somehow privileged over the others.

On the other hand, we have clear evidence that the vacuum geometry of the MSSM electroweak sector is not completely silent on the issue of R-parity and lepton number conservation. The most obvious evidence for this is that superpotentials which contain both elemental trilinears, $LLe + L\barH e$, always generate trivial vacuum moduli spaces. Thus, geometrical considerations seem to force the model-builder to take sides, between admitting $L\barH e$ as a fundamental trilinear, or $LLe$. The presence of these objects in the superpotential always lifts the corresponding GIO from the description of the vacuum moduli space. The difference between the resulting moduli spaces can then be identified in the nature of the syzygies between the trilinear operator with the opposite R-parity assignment and the bilinear GIOs. As these constraints descend from the $F_e$-term equations (arising from the F-terms associated with the charged lepton superfields), it is reasonable to assume that these lessons carry forward into a treatment of the full~MSSM.

In the absence of a right-handed neutrino sector, the inclusion of the term $L\barH e$ in the superpotential generates the Segre variety, which persists when a Higgs bilinear is added, but is trivialized by the R-parity odd bilinear $LH$ (Table~\ref{LbarHe-table}). In contrast, the inclusion of the term $LLe$ in the superpotential generates a slightly different vacuum manifold, which we have called the deformed Veronese variety. In this case inclusion of $H\barH$ trivializes the vacuum, and $LH$ reduces the variety to the conifold (Table~\ref{LLe-table}).  These become the starting points for further constraints which arise when the neutrino sector is introduced. Cases involving $L\barH e$ continue to be trivialized by the bilinear $LH$, but now the Dirac neutrino mass operator $LH\nu$ perturbs the vacuum manifold to the deformed Veronese. Severi varieties are only recovered when a Majorana mass term is included as well. 

However, the Veronese variety is not unique to the R-parity even fundamental trilinear. Both the case $W= L\barH e + LH\nu + \nu^2$ and $W= LLe + LH\nu + \nu^2$ give a Veronese surface. We can say that the non-trivial vacuum geometries clearly favor cases driven by the relations amongst the $LLe$ GIOs, and therefore a superpotential where this term is absent. But they are not restricted to these cases. More interesting is the ambiguity about the nature of lepton number assignments. The presence of a fundamental Higgs bilinear $H\barH$ is not necessary to generate a rich vacuum structure, and superpotentials containing $H\barH \nu$ and $H\barH (\nu+1)$ are equally capable of producing the Veronese variety.

It seems that the ultimate understanding of the geometrical nature of R-parity conservation may require a definite answer to the logically prior question of the geometrical implication of a conserved lepton number, carried by the neutrino field $\nu$. In such a world the set of GIOs to consider would change. Vacuum geometries would be different, even in the absence of superpotentials, because the relations between the new GIOs in a theory such as $SU(2)_L \times U(1)_Y \times U(1)_L$ would be different. We expect that such relations would be more complex, in that operators neutral under the combined gauge group would need to be larger in mass dimension, suggesting a larger elimination problem to compute. The defining polynomials of the vacuum varieties would likely have larger degree, for example. It would be of great interest to determine whether Severi varieties continue to arise in such circumstances, and whether a clearer preference for R-parity `even' superpotentials can be inferred.


\section*{Acknowledgements}

We thank James Gray and Mike Stillman for enjoyable collaborations on related themes.
YHH would like to thank the Science and Technology Facilities Council, UK, for grant ST/J00037X/1, the Chinese Ministry of Education, for a Chang-Jiang Chair Professorship at NanKai University as well as the City of Tian-Jin for a Qian-Ren Scholarship, as well as City University, London and Merton College, Oxford, for their enduring support.
VJ is supported by the DST/NRF South African Research Chairs Initiative.
CM is grateful to Damien Matti and EPFL for providing computing resources and is indebted to the National Institute for Theoretical Physics, the Mandelstam Institute for Theoretical Physics, and the University of the Witwatersrand for hospitality and financial support. BDN would like to thank the International Centre for Theoretical Physics (ICTP) in Trieste, Italy for hospitality during formative stages of this work. His work is supported by the U.S. National Science Foundation under grant PHY-0757959.







\end{document}